\documentstyle[aps,eqsecnum,twocolumn]{revtex}
\begin{document}
\draft
\title{Green Functions in Coordinate Space for Gauge Bosons
at Finite Temperature}
\author{H. Arthur Weldon}
\address{Department of Physics,
West Virginia University, Morgantown, West Virginia, 26506-6315}
\date{\today}
\maketitle
\begin{abstract}
The  thermal Green function, ${\cal
D}^{\mu\nu}(x)$, for free, massless gauge bosons  is computed exactly in
a variety of gauges  (Feynman, covariant, Coulomb, and Landshoff-Rebhan).
At large
temporal separations it falls exponentially. At large spatial
separations it falls like $T/r$.  In contrast, the
zero-temperature propagator falls quadratically in both regimes, being
proportional to
$1/(t^{2}-r^{2})$.
 \end{abstract}

\section{Introduction}

Although quantum field theories are formulated in four-dimensional coordinate
space, calculations are almost always performed in four-dimensional momentum
space. One major exception to the preference for momentum space
is non-equilibrium field theory.  Without equilibrium
there is no invariance under time-translation and often no
invariance under spatial translation. The dynamics of the density operator
$\varrho$ controls the evolution. For gauge bosons the time-ordered
propagator is
\begin{displaymath}
{\cal D}^{\mu\nu}_{11}(x,y)=-i{\rm Tr}\big(\varrho\,
T\big(A^{\mu}(x)A^{\nu}(y)\big)\big).
\end{displaymath}
In non-equilibrium the correlations depend not just on the separation $x\!-\!y$
but also on the effective age of the system $x\!+\!y$.
For situations in which  the
dynamical evolution depends rapidly on $x\!-\!y$ and slowly on $x\!+\!y$, it is
standard to Fourier transform from the separation $x\!-\!y$ to the conjugate
four-momentum $K$. The transformed propagator is then a function of $K$
and of
$x\!+\!y$. It has long been known \cite{1,2,3,4} how to use the
Schwinger-Dyson equation for the non-equilibrium propagator to
extract a kinetic equation for the  non-equilibrium
distribution function $f(K)$ that depends on $x\!+\!y$ . This approach has
been extended to modern field theories  containing gauge bosons particularly
for QCD\cite{5a,5,6,7,8}.

In ordinary vacuum field
theory, or zero-temperature field theory, the density operator is
$\varrho=|0\rangle\langle 0|$. Translation invariance is automatic and
calculations are almost always performed in momentum space.
In momentum space the time-ordered propagator for free gauge bosons in the
Feynman gauge is
$D_{11}^{\mu\nu}(K)=-g^{\mu\nu}/( K^{2}+i\eta)$.
This can easily  be Fourier transformed to give the coordinate-space propagator
\begin{equation}
{\cal D}_{11}^{\mu\nu}(x)\big|_{T=0}={-i\over 4\pi^{2}}{g^{\mu\nu}\over
x^{2}-i\epsilon}.
\label{1T=0}\end{equation}
Even though for fixed $t$  this falls like $1/r^{2}$ as $r\to\infty$, it
does contain the correct Coulomb potential, which  comes from the
light-cone singularity
$\delta(t\pm r)/r$.
At large time-like separations the $1/t^{2}$ behavior causes the electron
propagator to have a branch point at the electron mass shell instead of a
pole. The $1/x^{2}$ behavior often leads
to processes that are divergent both in the ultraviolet ($x\to 0$)
and in the infrared ($x\to\infty$).

A further consequence of
 Eq. (\ref{1T=0}) is that at zero temperature the quantum effects of massless
particles are quite complicated because the effects are as important
in space-like directions as in time-like directions.

\subsubsection*{Thermal Equilibrium}

For a field theory in an equilibrium heat bath that is uniform in space and
constant in time, there is a constant temperature $T$ throughout.
The density operator is given by $\varrho=e^{-H/T}/{\rm Tr}(e^{-H/T})$.
The time-ordered propagator  has  the structure \cite{b1,b2,b3}
\begin{equation}
{\cal D}_{11}^{\mu\nu}(x)=\theta(t){\cal D}^{\mu\nu}_{>}(x)
+\theta(-t){\cal D}_{<}^{\mu\nu}(x).
\end{equation}
The  thermal Wightman functions is defined by
\begin{equation}
{\cal D}^{\mu\nu}_{>}(x)=-i\sum_{n}e^{-\beta E_{n}}
{\langle n|A^{\mu}(x)A^{\nu}(0)|n\rangle\over
{\rm Tr}[e^{-\beta H}]},\label{Wight}\end{equation}
and ${\cal D}_{<}(x)={\cal D}_{>}(-x)$.
Appendix A summarizes how all other propagators (contour-ordered,
retarded, advanced) can be expressed in terms of ${\cal D}_{>}(x)$.

The thermal Wightman function satisfies two important conditions: Eq.
(\ref{R}) and (\ref{KMS}) below.  Both conditions come from the Heisenberg
relation
$A^{\mu}(t,\vec{r})=\exp(iHt)A^{\mu}(0,\vec{r})\exp(-iHt)$. This implies that
for complex time the field satisfies $A^{\mu}(t,\vec{r})^{\dagger}
=A^{\mu}(t^{\ast},\vec{r})$. Consequently the Wightman function
 enjoys the reflection property
\begin{equation}
[{\cal D}_{>}^{\mu\nu}(x)]^{\ast}=-{\cal D}_{>}^{\nu\mu}(-x^{*}).\label{R}
\end{equation}
Furthermore, by inserting a complete set of energy
eigenstates between the two field operators in Eq. (\ref{Wight}), it is simple
to show that at complex time  ${\cal
D}_{>}^{\mu\nu}(t,\vec{x})$ is analytic in the  strip
$-\beta\le{\rm Im}\, t\le 0$. A similar argument shows that the Wightman
function  must satisfy the Kubo-Martin-Schwinger  relation \cite{KMS}
\begin{equation}
{\cal D}_{>}^{\mu\nu}(t-i\beta,\vec{r})={\cal
D}_{>}^{\nu\mu}(-t,\vec{r}).\label{KMS}\end{equation}

In momentum space the Wightman function is given by
in terms of the
momentum-space spectral function $\rho^{\mu\nu}(K)$ by
\begin{equation}
D_{>}^{\mu\nu}(K)={-i\rho^{\mu\nu}(K)\over 1-e^{-\beta k_{0}}},
\label{start}
\end{equation}
which is derived in Appendix A.
The free spectral function  is independent of temperature
and only has support at
$k_{0}=\pm |\vec{k}|$. Thus only the on-shell particles contribute to the
free Wightman function.
The coordinate-space Wightman function is the Fourier transform
\begin{equation}
{\cal D}_{>}^{\mu\nu}(x)=\int {d^{4}K\over (2\pi)^{4}}e^{-iK\cdot x}
\;D_{>}^{\mu\nu}(K).
\label{spectral}\end{equation}
As shown in Appendix A,  the properties of the spectral function
guarantee that
${\cal D}_{>}^{\mu\nu}(x)$ will automatically satisfy the reflection
condition in Eq. (\ref{R}) and the KMS relation in Eq. (\ref{KMS}).

The remainder of the paper will perform the computation of Eq.
(\ref{spectral}) for free gauge bosons in various gauges.
Sec II deals entirely with the Feynman gauge. The complete result as a
function of arbitrary complex time is displayed in Eq. (\ref{2bGeneral}),
but a simpler form is the specialization to real time shown in Eq.
(\ref{B7}).  Sec III performs the calculation in a general covariant
gauge and leads to the results summarized in Eqns. (\ref{3Resulta}) and
(\ref{3Resultb}). Sec IV computes the Wightman function in the Coulomb gauge,
which is a bit more subtle, and the results are given in Eqns.
(\ref{Coul1}), (\ref{E1}), (\ref{Coul2}), and (\ref{E2}). Sec V contains a
 comparison of propagators with Bose-Einstein, Fermi-Dirac, and Boltzman
statistics. Appendix A provides some general formulas. Appendix B performs
the same calculation in the Landshoff-Rebhan quantization scheme
\cite{LR} in which the physical, transverse gauge fields are thermalized
but not the longitudinal and time-like components.

\section{${\cal D}^{\mu\nu}_{>}(\lowercase{x})$ in Feynman gauge}

It is easiest to compute the thermal Wightman function in the
Feynman gauge, in which the spectral function is proportional to the
constant metric tensor $g^{\mu\nu}$:
\begin{displaymath}
\rho^{\mu\nu}(K)=-
g^{\mu\nu}\,2\pi\epsilon(k_{0})\,\delta(K^{2}).\end{displaymath}
The thermal Wightman function has the form
\begin{equation}
{\cal D}_{>}^{\mu\nu}(x)=-g^{\mu\nu}{\cal D}_{>}(x),\end{equation}
where
\begin{displaymath}{\cal D}_{>}(x)=-i\int {d^{4}K\over
(2\pi)^{3}}e^{-iK\cdot x} {\epsilon(k_{0})\delta(K^{2})\over 1-e^{-\beta
k_{0}}}.
\end{displaymath}
The scalar function ${\cal D}_{>}(x)$ is, in addition, the thermal Wightman
function  for a spinless field of zero mass. To compute the integral, first
perform the integrals over
$k_{0}$ and over the angles of
$\vec{k}$ to obtain
\begin{equation}{\cal D}_{>}(x)={-1\over 8\pi^{2} r} \int_{0}^{\infty}dk
\,f_{k}(r,t),\label{2a}\end{equation}
where the $k$ dependence is contained in the function
\begin{displaymath}
f_{k}(t,r)\equiv\Big[e^{ikr}-e^{-ikr}\Big]
\Big[ {e^{-ikt}\over 1-e^{-\beta k} }-{e^{ikt}\over
1-e^{\beta k}}\Big].
\end{displaymath}
This function satisfies the conditions
\begin{eqnarray}
-f_{k}(t,r)^{\ast}&&=f_{k}(-t^{\ast},r)\nonumber\\
f_{k}(t-i\beta,r)&&=f_{k}(-t,r),\label{conditions}
\end{eqnarray}
which guarantee that ${\cal D}_{>}(x)$ will satisfy the reflection and KMS
conditions given in Eqs. (\ref{R}) and (\ref{KMS}).

\subsection{Unregularized ${\cal D}_{>}(x)$}

For any  real value of $t$, except for
 $t=\pm r$, the rapid oscillations of the integrand as $k\to\infty$ guarantee
convergence.  If $t$ has a negative imaginary part then convergence is improved
provided the negative imaginary part is smaller in magnitude than $\beta$.
Thus in the region
\begin{equation}
-\beta<{\rm Im}\,t\le 0\hskip0.5cm \&\hskip0.5cm t\neq \pm r
\label{2aRegion}\end{equation}
the integral is convergent without any regularization.
To perform the integration, it is convenient to use the fact that
$f_{k}(t,r)$ is an even function of $k$ in order to extend the  integration
range to the full  $k$ axis:
\begin{equation}{\cal D}_{>}(x)={-1\over 16\pi^{2} r}
\int_{-\infty}^{\infty}\!dk
\,f_{k}(r,t).\end{equation}
This can now be evaluated by contour integration over complex $k$.
The integrand $f_{k}(t,r)$ is finite at $k=0$. It has poles at $k=\pm i2\pi
nT$, for
integer $n\ge 1$.

 How the contour should be
closed depends on the relative size of ${\rm Re}\,t$ and $r$.
 For definiteness, let ${\rm Re}\,t>r$. Then
 $\exp(\pm ikt)$  is the determining factor.  For the  term containing
$\exp(-ikt)$ the contour should be closed in the lower-half of the complex
$k$ plane; for the term  $\exp(+ikt)$, the contour should be closed above.
 Cauchy's theorem gives the result as a sum of the residues
of the poles:
\begin{displaymath}
{\cal D}_{>}(x)={-iT\over 4\pi
r}\sum_{n=1}^{\infty}\big[(z_{+})^{n}-(z_{-})^{n}\big],
\end{displaymath}
 where the variables $z_{\pm}$ are defined by
\begin{equation}
z_{+}=e^{-2\pi T(t+r)}\hskip1cm z_{-}=e^{-2\pi T(t-r)}.
\label{2azpm}\end{equation}
This sum is only convergent in the region ${\rm Re}\,t>r$.
Performing the sum gives
\begin{equation}
{\cal D}_{>}(t,r)={-iT\over 4\pi r}\Big[{1\over e^{2\pi T(t+r)}-1}
-{1\over e^{2\pi T(t-r)}-1}\Big].\label{2aD>}\end{equation}
This result is valid for all complex $t$ satisfying Eq. (\ref{2aRegion})
and is analytic in this region. The poles at  $t=\pm r$ and at $t=\pm r-i\beta$
(limit points of the open region) will be shifted slightly
when the integration is regularized.
It is convenient to express ${\cal D}_{>}(x)$ in the alternate form
\begin{equation}
{\cal D}_{>}(t,r)\!=\!{-iT\over 8\pi r}\,\Big[\!\coth(\pi T(r\!+\!t))
\!+\!\coth(\pi T(r\!-\!t))\!\Big].\label{2acosh}\end{equation}
Note that this  is an even
function of $t$. The regularized ${\cal D}_{>}(x)$ will include the domain
$t=\pm r$ and will not be even under $t\to -t$.

\paragraph*{Deep time-like region:} For $|t|-r\gg 1/\pi T$,  the
asymptotic behavior is
\begin{equation}
{\cal D}_{>}(x)\rightarrow{iT\over 4\pi r}
\Big[e^{-2\pi T(|t|-r)}-e^{-2\pi T(|t|+r)}+\dots\Big].
\label{2time}
\end{equation}

\paragraph*{Deep space-like region:} For $r-|t|\gg 1/\pi T$, the asymptotic
behavior is not exponentially small.
\begin{equation}
{\cal D}_{>}(x)\rightarrow{-iT\over 4\pi r}
\!-\!{iT\over 4\pi r}\Big[e^{-2\pi T(r-|t|)}\!+\!e^{-2\pi
T(r+|t|)}\Big].\label{2space}
\end{equation}
The dominant term $T/r$ is present in all gauges and will be discussed
further.

\subsection{Regularized ${\cal D}_{>}(x)$}

The true Wightman function should be analytic in the closed region
\begin{equation}
-\beta\le{\rm Im}\,t\le 0,\label{2bRegion}\end{equation}
which includes the points $t=\pm r$ and $t=\pm r-i\beta$ at which
Eq.  (\ref{2aD>}) has poles.
 The problem is that at these values of $t$
 the integrand $f_{k}(t,r)\!\to\! 1$ without oscillation as $k\to
\infty$.
 The simplest way to regularize the integral in a way that will satisfy
Eqs. (\ref{R}) and (\ref{KMS}) is to define
\begin{equation}
{\cal D}_{>}(x)={-1\over 8\pi^{2}}\int_{0}^{\infty}dk\;f(t,r)\,e^{-\epsilon k},
\end{equation}
where $\epsilon$ is positive and real.  Now the range of integration cannot
be extended to negative $k$. However the integral can be evaluated using
the relation \cite{Gradshteyn}
\begin{displaymath}
{1\over T}\!\int_{0}^{\infty}\!dk\,{e^{-a_{1}k}-e^{-a_{2}k}\over e^{\beta k}-1}
=-\psi(1\!+\!a_{1}T)\!+\!\psi(1\!+\!a_{2}T),
\end{displaymath}
which holds whenever the real parts of $a_{1}$ and $a_{2}$ are positive.
Here
$\psi(z)=d\ln\Gamma(z)/dz$ . The result is
\begin{eqnarray}
{\cal D}_{>}(x)\!=\!&&{T\over 8\pi^{2}r}\Big[\psi[\epsilon T\!+\!iT(t\!-\!r)]
\!-\!\psi[\epsilon T\!+\!iT(t\!+\!r)]\\
&&-\psi[1+\epsilon T\!-\!iT(t\!-\!r)]
\!+\!\psi[1+\epsilon T\!-\!iT(t\!+\!r)]\Big].\label{psi1}\nonumber
\end{eqnarray}
This satisfies the reflection condition Eq. (\ref{R}), which interchanges the
first term with the second and interchanges the third term with the fourth. It
satisfies the KMS  condition (\ref{KMS}), which interchanges the first term
with
the third and the second term with the fourth.
Using $\psi(z)=\psi(1+z)-1/z$ this can be written so as to isolate the
zero-temperature contribution:
\begin{eqnarray}
{\cal D}_{>}(x)\!=\!
{T\over 8\pi^{2}r}&&\!
\bigg[\psi[1\!+\!T(\epsilon\!-\!i(r\!+\!t))]
\!-\!\psi[1\!+\!T(\epsilon\!+\!i(r\!+\!t))]\nonumber\\
+&&\psi[1\!+\!T(\epsilon\!-\!i(r\!-\!t))]
-\psi[1\!+\!T(\epsilon\!+\!i(r\!-\!t))]\bigg]\nonumber\\
-&&{i\over 4\pi^{2}}{1\over (r\!+\!t\!-\!i\epsilon)(r\!-\!t\!+\!i\epsilon)}
\label{2bGeneral}\end{eqnarray}
This is the complete and general result.
It is analytic in the closed region given in Eq. (\ref{2bRegion}).
The nearest poles to this region are the zero-temperature poles
just above the strip at $t=\pm r+i\epsilon$ and the the poles just below
the strip
at $t=\pm r -i(\beta+\epsilon)$ from the first and fourth term.

\paragraph*{Real time:} For real values of $t$ one can safely put
$\epsilon=0$ in all four $\psi$ functions since their poles are are at least
a distance
$\pm i\beta$ from the real $t$ axis.
When $\epsilon=0$ the psi functions have complex conjugate arguments and can be
simplified using \cite{AS}
\begin{displaymath}
\psi(1\!-\!iy)-\psi(1\!+\!iy)={i\over y}-i\pi\coth(\pi y).
\end{displaymath}
The Wightman function is
\begin{eqnarray}
{\cal D}_{>}(x)=&&{-iT\over 8\pi r}\Big[\coth[\pi T(r+t)+\coth[\pi T(r-t)\Big]
\nonumber\\
&&+{1\over 8\pi r}\big[\delta(r+t)-\delta(r-t)\big].\label{B7}
\end{eqnarray}
This, of course, agrees with Eq. (\ref{2acosh}) when $t\neq\pm r$.
Note that under $t\to -t$, the $\coth$ terms are symmetric but the
Dirac delta terms are antisymmetric.
The other Wightman function ${\cal D}_{<}(x)$ for real $t$ is
\begin{eqnarray}
{\cal D}_{<}(x)=&&{-iT\over 8\pi r}\Big[\coth[\pi T(r+t)+\coth[\pi T(r-t)]
\nonumber\\
&&+{1\over 8\pi r}\big[\delta(r-t)-\delta(r+t)\big].\label{B8}
\end{eqnarray}
Since Eqns. (\ref{B7}) and (\ref{B8}) hold only for real $t$, one cannot
pass from
one to the other by the KMS relation.

\def\sqr#1#2{{\vcenter{\vbox{\hrule height.#2pt
	\hbox{\vrule width.#2pt  height#1pt  \kern#1pt \vrule  width.#2pt}
	\hrule height.#2pt}}}}

\subsection{Time-ordered propagator}

The Wightman function satisfies the homogeneous equation
$\sqr66\,
{\cal D}_{>}(x)=0$. The time-ordered propagator is
\begin{displaymath}
{\cal D}_{11}(x)=\theta(t){\cal D}_{>}(x)+\theta(-t){\cal D}_{<}(x),
\end{displaymath}
in which ${\cal D}_{<}(t,r)={\cal D}_{>}(-t,r)$.
This is a true Green function in that it satisfies the inhomogeneous equation
$\sqr66\,
{\cal D}_{11}(x)=-\delta^{4}(x)$. Using Eq. (\ref{2bGeneral}) gives
 \begin{eqnarray}
{\cal D}_{11}(x)\!=\!{T\over
8\pi^{2}r}&&
\bigg[\psi[1\!+\!T(\epsilon\!-\!i(r\!+\!t))]
\!-\!\psi[1\!+\!T(\epsilon\!+\!i(r\!+\!t))]\nonumber\\
+&&\psi[1\!+\!T(\epsilon\!-\!i(r\!-\!t))]
-\psi[1\!+\!T(\epsilon\!+\!i(r\!-\!t))]\bigg]\nonumber\\
+{i\over 4\pi^{2}}&&\,{1\over
t^{2}-r^{2}-i\epsilon}.\label{2c}\end{eqnarray} This representation has the
nice feature that the zero-temperature limit is isolated in the last term.
The temperature-dependent terms are each annihilated by the  d'Alembertian
operator $\sqr66$ .

\subsection{Potential produced by a static charge}

In the limit $r\!\to\!\infty$ at fixed $t$,
 the Wightman function and  the time-ordered
propagator have the behavior
${\cal D}(x)\to -iT/(4\pi r)$. This contribution suggests that
at large distance the potential produced by a charge at rest would be
temperature-dependent. This inference  is incorrect as the following
calculation
demonstrates.

Let $J^{\mu}(x)$ be a classical current. In the Feynman gauge,
${\cal D}_{11}^{\mu\nu}(x)=-g^{\mu\nu}{\cal D}_{11}(x)$  and so the
classical vector potential is
\begin{eqnarray}
A^{\mu}_{\rm cl}(x)
=-\int d^{4}x'\,D_{11}(x-x')J^{\mu}(x').\nonumber\end{eqnarray}
For a point charge $Q$ at rest
$J^{0}(x')=Q\delta^{3}(\vec{r}^{\,\prime})$ and $\vec{J}(x')=0$. Thus the
three-vector potential vanishes and the scalar potential requires integrating
over the static charge density:
\begin{displaymath}
A^{0}_{\rm cl}(x)=- Q\int_{-\infty}^{\infty}\!dt'\;{\cal D}_{11}(t-t',r)
\end{displaymath}
It is convenient to use Eq. (\ref{2c}) for the Green function. At large complex
$t$ the combination of psi functions falls like $1/t^{2}$. Thus one can
integrate
over complex time by integrating over a contour $C$ that is closed in the
upper half-plane:
\begin{displaymath}
A^{0}_{\rm cl}(x)=- Q\oint_{C}\!dt'\;{\cal D}_{11}(t-t',r)
\end{displaymath}
The first and fourth $\psi$ function in Eq. (\ref{2c}) are analytic in the
upper-half of the complex $t$ plane and make no contribution.  The second and
third  $\psi$ functions   have poles in the upper half-plane at $t=\pm r
+i(N\beta
+\epsilon)$ for
$N=1,2,3\dots$. These poles all have the same residue (viz. $-i\beta$)  and
their
contributions to the potential cancel exactly.  Thus the entire potential
comes from the zero-temperature term:
\begin{displaymath}
A^{0}_{\rm cl}(x)=- {iQ\over 4\pi^{2}}\oint_{C}\!dt'\;
{1\over (t\!-\!t')^{2}\!-\!r^{2}\!-\!i\epsilon}.
\end{displaymath}
This is easily integrated and gives the usual Coulomb potential:
\begin{equation}A^{0}_{\rm cl}(x)={Q\over 4\pi r}.
\end{equation}
It is perhaps worth emphasizing that this not a large distance
approximation. A more difficult calculation, which does contain
temperature-dependence,  is the potential due to a point charge moving with
constant velocity, where
$J^{0}(x')=Q\delta^{3}(\vec{r}^{\,\prime}-\vec{v}\,t')$.

\section{${\cal D}^{\mu\nu}_{>}(\lowercase{x})$ in covariant
gauges}

The next case in which to compute the thermal Wightman functions
are the general covariant gauges. The tedious regularization performed in Sec
II will not be attempted. The result given in Eq. (\ref{3Resulta}) and
(\ref{3Resultb}) is therefore valid for real $t$  not on the light cone.

 In  a general covariant gauge the time-ordered propagator
at zero temperature is
\begin{displaymath}
D_{11}^{\mu\nu}(K)\big|_{T=0}={-g^{\mu\nu}\over K^{2}+i\epsilon}+(1-\xi){
K^{\mu}K^{\nu}\over ( K^{2}+i\epsilon)^{2}}.
\end{displaymath}
The  spectral function, extracted using Eq. (\ref{RA}), is
\begin{displaymath}
\rho^{\mu\nu}(K)=2\pi\epsilon(k_{0})[-g^{\mu\nu}+ (1-\xi)
K^{\mu}K^{\nu}{\partial
\over
\partial k^{2}}]
\delta(K^{2}).\end{displaymath}
When the thermal Wightman function ${\cal D}_{>}^{\mu\nu}(x)$ is expressed
in terms of the spectral function using Eq. (\ref{spectral}) the result is
\begin{equation}
{\cal D}_{>}^{\mu\nu}(x)=-g^{\mu\nu}{\cal D}_{>}(x)+(1-\xi)
{\partial^{2}\over \partial x_{\mu}\partial
x_{\nu}}d_{>}(x),\label{c1}\end{equation}
where ${\cal D}_{>}(x)$ with no
superscripts is the Wightman function from Sec II and the new function
$d_{>}(x)$  is
\begin{displaymath}
d_{>}(x)=i\int {d^{4}K\over (2\pi)^{3}}{e^{-iK\cdot x}\over
1-e^{-\beta k_{0}}}\epsilon(k_{0}){\partial\over\partial k^{2}}
\delta(K^{2}).    \end{displaymath}
It will be important later that ${\cal D}^{\mu\nu}_{>}(x)$ is not affected if a
constant is added to the value of $d_{>}(x)$.
The integration over $k_{0}$ and over the angles gives
\begin{equation}
d_{>}(x)={-i\over 16\pi^{2}} \int_{0}^{\infty}\!{dk\over
 k^{2}}\;g_{k}(t,r),\label{G>}
\end{equation}
where
\begin{displaymath}
g_{k}(t,r)=k\Big[e^{ikr}+e^{-ikr}\Big]\Big[{e^{-ikt}\over 1-e^{-\beta k}}
-{e^{ikt}\over 1-e^{\beta k}}\Big].\end{displaymath}
The function $g_{k}(t,r)$ equals $4T$ at $k=0$.
Consequently the integral in Eq. (\ref{G>}) does not converge at $k=0$.
However
the fact that $\partial g_{k}(t,r)/\partial t$ and $\partial
g_{k}(t,r)/\partial
r$ both vanish like $k^{2}$ as $k\to 0$ guarantees that $\partial
d_{>}(x)/\partial x_{\mu}$ is finite, which is all that is necessary for Eq.
(\ref{c1}). Since the behavior at $k=0$  is awkward,
 it is convenient to subtract a constant from Eq. (\ref{G>}) and
redefine $d_{>}(x)$ as
\begin{equation}
d_{>}(x)={-i\over 16\pi^{2}}\!\int_{0}^{\infty}\!{dk\over k^{2}}\,
\big[g_{k}(t,r)\!-\!g_{k}(0,0)\,{\mu^{2}\over k^{2}+\mu^{2}}\big].\label{G2}
\end{equation}
The integrand now has no singularity at $k=0$. A real parameter $\mu$
has been introduced so that the subtracted integration converges at $k=\infty$
for non-exceptional values of $t$.
Since it is not regulated as was done in Sec IIB, it does not
converge
 for $t=\pm r$ or for $t=\pm r -i\beta$.

It is simple to integrate Eq. (\ref{G2}). The integrand is even in $k$ and
thus the range may be extended to $-\infty$:
\begin{displaymath}
d_{>}(x)={-i\over 32\pi^{2}}\!\int_{-\infty}^{\infty}\!{dk\over k^{2}}\,
\big[g_{k}(t,r)\!-\!g_{k}(0,0)\,{\mu^{2}\over k^{2}+\mu^{2}}\big].
\end{displaymath}
The integral  can be evaluated by closing the contour in the  complex $k$ plane
and using Cauchy's theorem. How the contour is closed
depends upon
 the relative size of $t$ and $r$.  For example, if ${\rm Re}\,t>r$ the
contour for the terms containing $\exp(-ikt)$ should be closed  in the lower
half-plane; those containing $\exp(ikt)$ should be closed in the upper
half-plane. The integrand of (\ref{G2}) is not singular at
$k=0$. The poles at $k=\pm i2\pi nT$ for $n=1,2,3\dots$.
give the result
\begin{equation}
d_{>}(x)={-i\over 16\pi^{2}}\sum_{n=1}^{\infty}
\Big({(z_{+})^{n}\over n}+{(z_{-})^{n}\over n}\Big),\end{equation}
where the constant terms independent of $t$ and $r$ are omitted and
$z_{\pm}$ are the same as in Eq. (\ref{2azpm}).
The summation  displayed is only convergent if the magnitudes of $z_{\pm}$
are less than one: i.e.   ${\rm
Re}\,t>r$. Performing the sum in this region gives
\begin{equation}
d_{>}(x)={i\over 16\pi^{2}}\Big[\ln(1-z_{+})
+\ln(1-z_{-})\Big].\end{equation}
This result can now be extended to ${\rm Re}\,t<r$. Since it was not
regularized, it is not valid precisely at
$z_{+}=1$ or at
$z_{-}=1$. The result is not symmetric under $t\to -t$, but it can be rewritten
as
\begin{eqnarray}
d_{>}(x)=&&{i\over 16\pi^{2}}\ln\Big[\sinh(\pi T(r\!+\!t))
\sinh(\pi T(r\!-\!t))\Big]\nonumber\\
&&-{iT\over 8\pi}t+{\rm constant}.\label{3Resulta}
\end{eqnarray}
The term linear in $t$ and the constant do not have second derivatives and thus
do not affect the value of $D_{>}^{\mu\nu}(x)$. The function $d_{>}(x)$ is
related
to the scalar Wightman function by
$\sqr66\,d_{>}(x)= {\cal D}_{>}(x)$.  Therefore Eq. (\ref{c1}) can be
summarized by
\begin{equation}
{\cal D}_{>}^{\mu\nu}(x)=\Big(-g^{\mu\nu}\sqr66+(1-\xi)\,\partial^{\mu}
\partial^{\nu}\Big)\,d_{>}(x).\label{3Resultb}
\end{equation}
The form of this in various limits will now be examined.

\paragraph*{Deep time-like region:} When $|t|-r\gg 1/\pi T$ the
asymptotic behavior of Eq. (\ref{3Resulta}) is
\begin{displaymath}
d_{>}(x)\to {-i\over 16\pi^{2}}\Big[e^{-2\pi T(|t|-r)}+e^{-2\pi
T(|t|+r)}\Big],
\end{displaymath}
with an irrelevant term linear in $t$ omitted.
Consequently all components of ${\cal D}_{>}^{\mu\nu}(x)$ fall
exponentially as was the case in Feynman gauge.

\paragraph*{Deep space-like region:} If $r-|t|\gg 1/\pi T$, then
Eq. (\ref{3Resulta}) has the behavior
\begin{displaymath}
d_{>}(x)\to {iTr\over 8\pi}\!-\!{i\over 16\pi^{2}}\Big[e^{-2\pi
t(r-|t|)} +e^{-2\pi T(r+|t|)}\Big],
\end{displaymath}
omitting additive constants. The asymptotic behavior of
 ${\cal D}^{00}_{>}(x)$ is
\begin{mathletters}\begin{equation}
{\cal D}_{>}^{00}(x)\to  {iT\over 4\pi r}+{\cal O}(e^{-2\pi Tr}).
\end{equation}
The leading term, being independent of $\xi$, coincides with the Feynman
gauge result. It was only this leading term that contributed to the
calculation of the Coulomb potential in Sec II D.
Of the remaining components,  ${\cal D}_{>}^{0j}(x)$ is exponentially small:
\begin{equation}
{\cal D}_{>}^{0j}(x)=(1-\xi)\,{\partial^{2}G_{>}(x)\over \partial
t\partial x_{j}}\to {\cal O}(e^{-2\pi Tr}).
\end{equation}
For the spatial components of the propagator, the term in $d_{>}(x)$ that is
linear in
$r$ contributes a term from
\begin{displaymath}
{\partial^{2}\over\partial x_{i}\partial x_{j}}r={1\over
r}(\delta^{ij}
-\hat{x}^{i}\hat{x}^{j}).
\end{displaymath}
The asymptotic form of the spatial propagator is
\begin{equation}
{\cal D}_{>}^{ij}(x)\to {iT\over 8\pi r}\big[\!-\!(\xi\!+\!1)\delta^{ij}
\!+\!(\xi\!-\!1)
\hat{x}^{i}\hat{x}^{j}\big]\!+\!{\cal O}(e^{-2\pi Tr})
\end{equation}\end{mathletters}

\paragraph*{Zero-temperature limit:} At zero temperature, $d_{>}(x)$ is
proportional to $\ln[r^{2}-t^{2}]$ and  Eq.
(\ref{3Resultb}) gives
\begin{equation}
{\cal D}_{>}^{\mu\nu}(x)\big|_{T=0}=-(\xi\!+\!1){i\over
8\pi^{2}}{g^{\mu\nu}\over x^{2}}
+(\xi\!-\!1){i\over 4\pi^{2}}{x^{\mu}x^{\nu}\over (x^{2})^{2}}.
\label{3T=0}\end{equation}
The combination $x_{\mu}x_{\nu}{\cal
D}_{>}^{\mu\nu}(x)\big|_{T=0}=i(\xi\!-\!3)/8\pi^{2}$ and thus vanishes in the
Yennie gauge.

\section{${\cal D}^{\lowercase{i\,j}}_{>}(\lowercase{x})$ in Coulomb gauge}

In the Coulomb gauge the time-like component of the gauge
potential is instantaneous and does not propagate in time.
Consequently
\begin{displaymath}
{\cal D}^{00}(x)={\delta(t)\over4\pi r},\end{displaymath}
and there are no thermal corrections.
 For transverse gauge bosons the zero-temperature propagator is
\begin{displaymath}
D_{F}^{ij}(K)=\Big(\delta^{ij}-{k^{i}k^{j}\over k^{2}}\Big)
{1\over K^{2}+i\epsilon}.\end{displaymath}
and the transverse spectral function is
\begin{displaymath}
\rho^{ij}(K)=2\pi\epsilon(k_{0})\delta(K^{2})
\big[\delta^{ij}-\hat{k}^{i}\hat{k}^{j}\big].
\end{displaymath}
Using Eq. (\ref{spectral}), the two-point function can be written
\begin{equation}
{\cal D}_{>}^{ij}(x)=\delta^{ij}{\cal D}_{>}(x)+\nabla^{i}
\nabla^{j}\,H_{>}(x),\label{3a}\end{equation}
where the first term, ${\cal D}_{>}(x)$, is the same as
 in Sec. II and the new function is
\begin{displaymath}
H_{>}(x)=-i\int {d^{4}K\over (2\pi)^{3}}
{e^{-iK\cdot x}\over 1-e^{-\beta k_{0}}}
{\epsilon(k_{0})\delta(K^{2})\over k^{2}}.
\end{displaymath}
By construction, ${\cal D}_{>}(x)=-\nabla^{2}H_{>}(x)$.

To compute $H_{>}(x)$ the first step is to integrate
 over $k_{0}$ and the angles to obtain
\begin{equation}
H_{>}(x)={-1\over 8\pi^{2}}\int_{0}^{\infty}\!{dk\over k^{2}}
\;h_{k}(t,r),\label{H>}
\end{equation}
where
\begin{displaymath}
h_{k}(t,r)={1\over r}\Big[e^{ikr}-e^{-ikr}\Big]
\Big[{e^{-ikt}\over 1-e^{-\beta k}}-{e^{ikt}\over 1-e^{\beta
k}}\Big].
\end{displaymath}
For later reference note that $H_{>}(x)$ is analytic for complex time in the
open region $-\beta<{\rm Im}\,t<0$.

The function $h_{k}(t,r)$ equals $4iT$ at $k=0$. Consequently
the integral over $k$ does not converge at $k=0$. However
$\partial H_{>}(x)/\partial r$ is convergent, which is all that is
necessary for Eq. (\ref{3a}).
To improve the $k=0$ behavior it is convenient to subtract an
$x$-independent constant from Eq. (\ref{H>}) and redefine
$H_{>}(x)$ as
\begin{displaymath}
H_{>}(x)={-1\over 8\pi^{2}}\int_{0}^{\infty}\!{dk\over k^{2}}
\;\big[h_{k}(t,r)-h_{k}(0,0){\mu^{2}\over
k^{2}+\mu^{2}}\big],
\end{displaymath}
where $\mu$ is an unimportant mass parameter.
The integral is now convergent at $k=0$ and at $k=\infty$.
Since the integrand  is even in $k$, the range
may be extended over negative $k$:
\begin{equation}
H_{>}(x)=\!{-1\over 16\pi^{2}}\!\int_{-\infty}^{\infty}\!{dk\over
k^{2}}\;\big[h_{k}(t,r)\!-\!h_{k}(0,0){\mu^{2}\over
k^{2}+\mu^{2}}\big].\label{H>2}
\end{equation}
This will now be calculated explicitly for real $t$, without the
regularization that was performed in Sec II.

\paragraph*{Time-like region:}
When $t>r$ the contour for the  $e^{-ikt}$ contribution may  may be
closed in the lower half-plane and that for the  $e^{ikt}$ contribution
 closed in the upper half-plane. Although the full integrand has no
poles at $k=0$, each of these pieces has a double pole at $k=0$.
The more important contribution comes from the simple poles at $k=\pm i2\pi
nT$ for $n\ge 1$. The result is
\begin{equation}
H_{>}(x)={-i\over 16\pi^{3}rT}
\Big[{\rm Li}_{2}(z_{-})-{\rm
Li}_{2}(z_{+})\Big]\!
+\!{i\over 4\pi}Tt\!-\!{1\over 8\pi}.\label{Htime}\end{equation}
The last two terms in come from the double pole at $k=0$ and do not
contribute to  $\partial H_{>}(x)/\partial r$. The variables
$z_{\pm}$ are given in Eq. (\ref{2azpm}) and ${\rm Li}_{2}$ is the
dilogarithm function, which for $z\le 1$ has the series expansion
\cite{AS,Lewin}
\begin{equation}
{\rm Li}_{2}(z)\equiv\sum_{n=1}^{\infty}{z^{n}\over n^{2}}.
\label{Li1}\end{equation}
When $H_{>}(x)$ is substituted into Eq. (\ref{3a}) the result is
\begin{equation}
{\cal D}_{>}^{ij}(x)=(\delta^{ij}\!-\!\hat{x}^{i}\hat{x}^{j}){\cal D}_{>}(x)
+(\delta^{ij}\!-\!3\hat{x}^{i}\hat{x}^{j})E(x)
\label{Coul1}\end{equation}
where the new function $E(x)$ is
\begin{eqnarray}
E(x)=&&{i\over
8\pi^{2}r^{2}}\big[\ln(1-z_{-})+\ln(1-z_{+})\big]\nonumber\\
&&+{i\over 16\pi^{3}r^{3}T}\big[{\rm Li}_{2}(z_{-})-{\rm Li}_{2}(z_{+})\big].
\label{E1}\end{eqnarray}

In the deep time-like region, $\pi T(|t|-r)\gg 1$, both $z_{+}$ and $z_{-}$
are exponentially small and so is $E(x)$. Since ${\cal D}_{>}(x)$ vanishes
exponentially as shown in Eq. (\ref{2time}), this means that all terms in
the Wightman function
${\cal D}_{>}^{ij}(x)$ fall exponentially in the deep time-like region.

\paragraph*{Space-like region:} When $r>|t|$ it is best to return to the
defining integral Eq. (\ref{H>2}).
In that integration, the contour for the
$e^{ikr}$ contribution may  may be closed in the upper half-plane and
the contour for the  $e^{-ikr}$ contribution
 closed in the lower half-plane. As previously noted the full integrand
has no poles at $k=0$. However each of these pieces has a triple pole at
$k=0$. The simple poles at $k=\pm i 2\pi nT$ give infinite series:
\begin{eqnarray}
H_{>}(x)=&&{i\over 16\pi^{3}rT}\Big[{\rm Li}_{2}\big({1\over
z_{-}}\big)+{\rm Li}_{2}(z_{+})\Big]\nonumber\\
&&+{i\over 8\pi}\big[T\big(r+{t^{2}\over r}\big)+i{t\over r}-{1\over
6rT}\big].\label{Hspace}
\end{eqnarray}
The last term proportional to $1/rT$ is particularly unusual.

As mentioned previously, the original integral is analytic in complex
time provided $-\beta<{\rm Im}\,t<0$.
However it is not obvious that Eq. (\ref{Hspace}) is the analytic continuation
of Eq. (\ref{Htime}), and this provides an important check.  In
both results the arguments of the dilogarithm functions  are smaller than one.
For
$|z|<1$ the dilogarithm has the integral  representation
\cite{AS,Lewin}
\begin{displaymath}
{\rm Li}_{2}(z)=\int_{0}^{1}\!ds\;(\ln s){1\over s-z^{-1}}.\hskip1cm
(|z|<1)
\end{displaymath}
When the time is made complex by $t\to t-i\sigma$ then $z_{\pm}=e^{i2\pi
T\sigma}|z_{\pm}|$. Thus if $0<\sigma<\beta/2$ then $z_{\pm}$ has a
positive imaginary part.
For $|z|>1$ the analytic continuation is
\begin{displaymath}
{\rm Li}_{2}(z)\!=\!\int_{0}^{1}\!ds\,(\ln s){{\cal P}\over
s-z^{-1}}+i\pi\ln(z)\hskip.5cm\cases{|z|>1\cr {\rm Arg}(z)>0}.
\end{displaymath}
Omission of the imaginary part, as is usually done in textbooks,
would prevent the function from being analytic.
From the integral representation it follows that for $|z|>1$ and
${\rm Arg}(z)>0$ the dilogarithm satisfies
\begin{equation}
{\rm Li}_{2}(z)={\pi^{2}\over 3}-{1\over 2}(\ln z)^{2}
-{\rm Li}_{2}\Big({1\over z}\Big)+i\pi\ln(z).\end{equation}
Using this relation one can analytically continue Eq. (\ref{Htime}) from the
time-like region where $z_{-}<1$ to the space-like region where $z_{-}>1$ and
obtain Eq. (\ref{Hspace}). Thus the two results agree despite their appearance.

The Wightman function in the space-like region results from substituting  Eq.
(\ref{Hspace}) into Eq. (\ref{3a}):
\begin{equation}
{\cal D}_{>}^{ij}(x)=(\delta^{ij}\!-\!\hat{x}^{i}\hat{x}^{j})
{\cal D}_{>}(x) +(\delta^{ij}-3\hat{x}^{i}\hat{x}^{j})E(x),\label{Coul2}
\end{equation}
where $E(x)$ is now given by
\begin{eqnarray}
E(x)=&&{iT\over 8\pi r}-{i\over 8\pi r^{3}}(Tt^{2}+it-{1\over 6T})\nonumber\\
&&+{i\over
8\pi^{2}r^{2}}\big[\ln(1-{1\over z_{-}})+\ln(1-z_{+})\big]\nonumber\\
&&-{i\over 16\pi^{3}r^{3}T}\big[{\rm Li}_{2}({1\over z_{-}})+{\rm
Li}_{2}(z_{+})\big].\label{E2}
\end{eqnarray}
Naturally this is the analytic continuation of Eq. (\ref{E1}).
It is easy to check that $\nabla_{i}{\cal D}^{ij}_{>}(x)=0$.

In the deep space-like region defined by $\pi T(r-|t|)\gg 1$, both
$1/z_{-}$ and $z_{+}$ are exponentially small. The asymptotic behavior of the
Wightman function is
\begin{eqnarray}
{\cal D}_{>}^{ij}(x)\to &&(\delta^{ij}\!+\!\hat{x}^{i}\hat{x}^{j})
\Big({-iT\over 8\pi r}\Big)\\
+&&(\delta^{ij}\!-\!3\hat{x}^{i}\hat{x}^{j})
\Big({-i\over 8\pi r^{3}}\Big)
\big(Tt^{2}+it-{1\over 6T}\big).\nonumber
\end{eqnarray}
 Despite its
complicated appearance this still satisfies the transverse condition
$\nabla_{i}{\cal D}^{ij}_{>}(x)=0$.
The result is  more complicated than in covariant gauges, which
behaved as $T/r$ with exponentially small corrections.
Here the corrections are powers of $r$: the second line being  order
$1/r^{3}$ at fixed $t$  but  order $1/r$ if the ratio $t/r$ is fixed
as $r\!\to\!\infty$.

\paragraph*{Zero-temperature limit:}
It is straightforward to evaluate the $T=0$ limit Coulomb-gauge Wightman
function. As $T\to 0$ the arguments  of the dilogarithm functions
in Eq. (\ref{E1}) approach
unity: $z\to 1$.
A useful expansion in this region is \cite{AS,Lewin}
\begin{equation}
{\rm Li}_{2}(z)={\pi^{2}\over 6}-\ln(z)\ln(1-z)
-\sum_{n=1}^{\infty}{(1-z)^{n}\over n^{2}}.\end{equation}
The Wightman function becomes
\begin{eqnarray}
{\cal D}^{ij}_{>}\Big|_{T=0}
\!=&&\!(\delta^{ij}\!-\!\hat{x}^{i}\hat{x}^{j}){i\over 4\pi^{2}
(t^{2}-r^{2})}\nonumber\\
+&&(\delta^{ij}\!-\!3\hat{x}^{i}\hat{x}^{j}){i\over 8\pi^{2}}
\Big({2\over r^{2}}-{t\over r^{3}}\ln\Big[{t+r\over t-r}\Big]
\Big).
\end{eqnarray}
Naturally this agrees with the direct Fourier transform of the
zero-temperature propagator.

\section{Discussion}

The  thermal Wightman  function for free gauge
bosons has been computed in various gauges.
Knowing the Wightman function  is the same as knowing the time-ordered
propagators  as shown in Appendix A. (The free retarded and
advanced propagators are unchanged by the temperature since the free-field
commutator is a c-number and thermal averaging does not change the c-number.)

The rather surprising result is  that  the
large-distance effects of massless gauge bosons   are  simpler
at $T\!>\!0$  than in vacuum. In vacuum the free Green functions
fall like $1/(t^{2}-r^{2}).$  The thermal Green functions in covariant
gauges and in the Coulomb gauge are exponentially small at large time-like
separations. At large space-like separations the leading behavior is $T/r$,
with exponentially small corrections in covariant gauges
and power-law corrections in the Coulomb gauge. Appendix B computes the
Wightman functions for the Landshoff-Rebhan approach to thermalization
\cite{LR}.

Since the calculations presented all involve the Bose-Einstein distribution
function,  it is not apparent how much  the asymptotic behavior depends on
that particular distribution function.  The three examples below with
different statistics  will show that the results obtained
in Secs II-IV  are a consequence of quantum Bose
statistics.

\paragraph*{Bose-Einstein Statistics:}
For Bose-Einstein statistics the simplest case is that of a massless
scalar boson. The time-ordered propagator in momentum-space is
\begin{equation}
D_{11}(K)={1\over K^{2}+i\eta}-{2\pi i\,\delta(K^{2})\over
e^{\beta |k_{0}|}-1}.\label{Bose1}
\end{equation}
The Fourier transform of this has been computed in the Feynman gauge
discussion Sec II.  For real times not on the light cone, a useful
representation is that of Eq. (\ref{2acosh}):
\begin{equation}
{\cal D}_{>}(x)={-iT\over 4\pi r}
\Big[\coth(\pi T(r\!+\!t))\!+\!\coth(\pi T(r\!-\!t))\Big].
\label{Bose2}
\end{equation}
The asymptotic behavior as $r\to\infty$ can be understood by considering the
propagator in the Matsubara formulation, which has discrete frequencies
$\omega_{n}=2n\pi T$ because of the periodicity condition in imaginary time
\cite{b1,b2,b3}.
As observed by Linde \cite{ADL,GPY} the  $n=0$ modes produce the dominant
behavior in the deep space-like region.
The $n=0$ mode contribution is  exactly $-iT/(4\pi r)$.
However  the large time-like behavior does not come from a single mode in  the
Matsubara formalism, but requires summing a series and analytically
continuing  from complex time to real time. The complete
result is the same as Eq. (\ref{Bose2}).

\paragraph*{Fermi-Dirac Statistics:}
A convenient way to display the role of statistics is to change the
distribution function in Eq. (\ref{Bose1}) from Bose-Einstein to
Fermi-Dirac:
\begin{equation}
D_{11}(K)={1\over K^{2}+i\eta}+{2\pi i\,\delta(K^{2})\over
e^{\beta |k_{0}|}+1}.\label{Fermi1}
\end{equation}
This differs from Eq. (\ref{Bose1}) in that the $-1$ in the denominator has
been changed to a $+1$ and also the overall sign of the second term has been
changed. This is done so that the propagator for a
massless spin 1/2 fermion is the product of Eq. (\ref{Fermi1})
with $\gamma^{\mu}K_{\mu}$ \cite{b1,b2,b3}.  The Fourier transform of
Eq. (\ref{Fermi1}) can be easily computed along the lines indicated in Sec II.
For real times not on the light cone, the result is
\begin{equation}
{\cal D}_{>}(x)\!=\!{-iT\over 4\pi r}
\Big[{1\over\sinh(\pi T(r\!+\!t))}+{1\over\sinh(\pi T(r\!-\!t))}
\Big].
\end{equation}
At zero temperature this is the same as for Bose-Einstein statistics.
At large time-like separations it falls exponentially as in the Bose-Einstein
case.  At large space-like separations it also falls exponentially
in  contrast to Eq. (\ref{Bose2}). The space-like behavior can be
understood in the Matsubara formalism, since  the fermion
frequencies $\omega_{n}=(2n+1)\pi T$ can never vanish.

\paragraph*{Classical Statistics:}
Classical Boltzman statistics gives  a very different result.
In the classical limit the thermal distribution function in Eq. (\ref{Bose1})
is no longer singular at $k_{0}=0$:
\begin{equation}
D_{11}(K)={1\over K^{2}+i\eta}-{2\pi i\,\delta(K^{2})\over
e^{\beta |k_{0}|}}.\label{Class1}
\end{equation}
The Fourier transform of this gives
\begin{displaymath}
{\cal D}_{>}(x)={-i\over 4\pi^{2}}
\Big[{1\over r^{2}\!-\!(t\!-\!i\epsilon)^{2}}
\!+\!{1\over r^{2}\!-\!(t\!-\!i\beta)^{2}}
\!+\!{1\over r^{2}\!-\!(t\!+\!i\beta)^{2}}\Big]
\end{displaymath}
At zero temperature this is the same as the Bose or Fermi case.
However at $T\neq 0$ it is quite different. For large time-like separations
it falls like $1/t^{2}$. For large space-like separations it also falls like
$1/r^{2}$. Thus, for classical statistics the temperature does not
substantially change the asymptotic behavior.

 Subsequent publications will explore the physical consequences
of the space-time behavior. It should be possible to understand the
hard thermal loop approximation \cite{BP} directly in coordinate space.
In that approximation the high temperature corrections come entirely from
one-loop diagrams. The number of external lines determines the
number of internal propagators. The important effects come from thermal
corrections on one internal propagator with  all others kept at their
zero-temperature value. A coordinate-space analysis should explain why
only one loop diagrams are important  and why only one propagtor in the
loop enjoys thermal corrections. In addition, since non-equilibrium
processes almost demand a coordinant space treatment, it may be possible to
deduce  an extension of
the hard loop approximation to non-equilibrium processes.

\acknowledgements

This work was supported in part by the U.S. National Science Foundation under
grant PHY-9900609.

\appendix

\section{General structure}

This appendix  summarizes some standard properties of gauge boson propagators
at finite temperature \cite{b1,b2,b3}.
The basic thermal Wightman function is
\begin{equation}
{\cal D}^{\mu\nu}_{>}(x)=-i\sum_{n}e^{-\beta E_{n}}
{\langle n|A^{\mu}(x)A^{\nu}(0)|n\rangle\over
{\rm Tr}[e^{-\beta H}]}.\label{A1}\end{equation}
It is customary to introduce a special notation ${\cal D}_{<}(x)$ for the
function with inverted space-time arguments:
 \begin{equation}
 {\cal D}_{<}^{\mu\nu}(x)\equiv{\cal
D}_{>}^{\nu\mu}(-x).\label{defD<}\end{equation}
In terms of the field operators this means that
\begin{equation}
{\cal D}^{\mu\nu}_{<}(x)=-i\sum_{n}e^{-\beta E_{n}}
{\langle n|A^{\nu}(0)A^{\mu}(x)|n\rangle\over
{\rm Tr}[e^{-\beta H}]}.
\end{equation}
The retarded and advanced propagators are given by
\begin{eqnarray}
{\cal D}_{R}^{\mu\nu}(x)=\theta(t)&&\Big[{\cal D}^{\mu\nu}_{>}(x)
-{\cal D}^{\mu\nu}_{<}(x)\Big]\nonumber\\
{\cal D}_{A}^{\mu\nu}(x)=\theta(-t)&&\Big[{\cal D}^{\mu\nu}_{<}(x)
-{\cal D}^{\mu\nu}_{>}(x)\Big].\nonumber
\end{eqnarray}
For any parameter $\sigma$ in the range $0\le\sigma\le\beta$, the
 four parts of the contour-ordered propagator are
\begin{eqnarray}
{\cal D}_{11}^{\mu\nu}(x)=&&
\theta(t){\cal D}_{>}^{\mu\nu}(x)+\theta(-t){\cal
D}_{<}^{\mu\nu}(x)\nonumber \\
{\cal D}_{12}^{\mu\nu}(x) =
&& {\cal D}_{<}^{\mu\nu}(t+i\sigma,\vec{x})\nonumber\\
{\cal D}_{21}^{\mu\nu}(x)=
&&{\cal D}_{>}^{\mu\nu}(t-i\sigma,\vec{x})\nonumber\\
{\cal D}_{22}^{\mu\nu}(x)=
&&\theta(t){\cal D}_{<}^{\mu\nu}(x)+\theta(-t){\cal
D}_{>}^{\mu\nu}(x).\nonumber\end{eqnarray}
With respect to real time, the first of these is time-ordered; the fourth is
anti-time-ordered.

The spectral function is the thermal average of the commutator
\begin{displaymath}
\rho^{\mu\nu}(x)=\sum_{n}e^{-\beta E_{n}}
{\langle n|\big[A^{\mu}(x),A^{\nu}(0)\big]|n\rangle\over
{\rm Tr}[e^{-\beta H}]}.
\end{displaymath}
Its Fourier transform $\rho^{\mu\nu}(K)$ has the properties
\begin{eqnarray}
\rho^{\mu\nu}(-K)=&&-\rho^{\nu\mu}(K)\\
\rho^{\mu\nu}(K)^{\ast}=&&\rho^{\nu\mu}(K).\end{eqnarray}
The Fourier transform of Eq. (\ref{A1}) is
\begin{equation}
D_{>}^{\mu\nu}(K)=-i
{\rho^{\mu\nu}(K)\over1- e^{-\beta k_{0}}}.\end{equation}
This is the starting point in Eq. (\ref{start}).
The spectral function is most easily obtained from
 the propagator in momentum space by the relation
\begin{equation}
\rho^{\mu\nu}(K)=i\big[D^{\mu\nu}_{R}(K)-D^{\mu\nu}_{A}(K)\big].
\label{RA}
\end{equation}

\section{Landshoff-Rebhan propagator}

Landshoff and Rebhan have advocated heating only the two physical (i.e.
spatially transverse) components of the gauge potential even in a
general covariant gauge \cite{LR}. The procedure is to begin with the $T=0$
propagator in a general  covariant gauge,
\begin{displaymath}
D_{11}^{\mu\nu}(K)={-g^{\mu\nu}\over K^{2}+i\epsilon}+(1-\xi){
K^{\mu}K^{\nu}\over ( K^{2}+i\epsilon)^{2}}.
\end{displaymath}
As explained in \cite{LR},
the $D^{00}(K)$ and $D^{0j}(K)$ components are unaffected by temperature and
therefore the corresponding Wightman functions in space-time may be
read off from Eq. (\ref{3T=0}):
\begin{eqnarray}
{\cal D}_{>}^{00}(x)=&&{i\over 8\pi^{2}}
\;{(\xi-3)t^{2}+(\xi+1)r^{2}\over (x^{2}-i\eta)^{2}}\label{a}\\
{\cal D}_{>}^{0j}(x)=&&{i\over 4\pi^{2}}(\xi-1){tx^{j}\over
(x^{2}-i\eta)^{2}}.\label{b}
\end{eqnarray}
To separate  the physical components of $\vec{A}$,
rearrange the spatial components of the $T\!=\!0$ propagator as
\begin{displaymath}
D^{ij}_{11}(K)={\delta^{ij}-k^{i}k^{j}/k^{2}
\over K^{2}+i\epsilon}+{(k_{0}^{2}-\xi k^{2})k^{i}k^{j}/k^{2}\over
(K^{2}+i\epsilon)^{2}}.\end{displaymath}
The first term describes physical, transversely polarized, particles and is
heated. The second term comes from  gauge-dependent, longitudinally
polarized particles  and is not heated. From the previous results the
finite-temperature two point function is
\begin{equation}
{\cal D}_{>}^{ij}(x)={\cal D}_{>\rm Coul}^{ij}(x)+F_{>}^{ij}(x)
\label{c}\end{equation}
where the temperature-dependent Coulomb part is given by Eqs.
(\ref{Coul1}), (\ref{E1}) in the time-like region and by
Eqns. (\ref{Coul2}), (\ref{E2}) in the space-like region. The new
temperature-independent contribution $F^{ij}_{>}(x)$ is
\begin{eqnarray}
F^{ij}_{>}(x)=&&(\delta^{ij}\!-\!\hat{x}^{i}\hat{x}^{j}){i\over
16\pi^{2}} {\xi+1\over t^{2}-r^{2}}
+\hat{x}^{i}\hat{x}^{j}{i\over 4\pi^{2}}{t^{2}\over (t^{2}-r^{2})^{2}}
\nonumber\\
+&&(\delta^{ij}\!-\!3\hat{x}^{i}\hat{x}^{j}){i\over 16\pi^{2}}
\Big({\xi\!-\!3\over t^{2}\!-\!r^{2}}\!-\!{4\over r^{2}}
\!+\!{2t\over r^{3}}\ln\Big[{t+r\over t-r}\Big]\Big).\nonumber
\end{eqnarray}

The Landshoff-Rebhan propagator clearly has a very complicated asymptotic
behavior. In the deep space-like region, $r\to\infty$, Eqs. (\ref{a})
fall like
$1/r^{2}$, Eq. (\ref{b}) falls like $1/r^{3}$,  whereas the Coulomb part of
Eq. (\ref{c}) still falls like $1/r$. In the deep time-like region,
$t\to\infty$,  Eqns. (\ref{a}), (\ref{b}), and (\ref{c}) are all dominated
by the zero-temperature contributions and  fall like
$1/t^{2}$ rather than exponentially.

\end{document}